\documentstyle[emulateapj]{article}

\slugcomment{Submitted to the Astrophysical Journal }

\lefthead{Fesen et al.}
\righthead{{\it HST\/} Images and Spectra of SN~1885}

\newcommand{\be}{\begin{equation}}
\newcommand{\ee}{\end{equation}}
\newcommand{\ba}{\begin{eqnarray}}
\newcommand{\ea}{\end{eqnarray}}

\newcommand{\Ca}{{\rm Ca}}
\newcommand{\CaI}{{\rm Ca I}}
\newcommand{\CaII}{{\rm Ca II}}
\newcommand{\Fe}{{\rm Fe}}
\newcommand{\FeI}{{\rm Fe I}}
\newcommand{\FeII}{{\rm Fe II}}
\newcommand{\Msun}{\mbox{${\rm M}_{\sun}$}}
\newcommand{\arcsecs}{{\rm arcsec}}
\newcommand{\cm}{{\rm cm}}
\newcommand{\erg}{{\rm erg}}
\newcommand{\eV}{{\rm eV}}
\newcommand{\kms}{{\rm km}\,{\rm s}^{-1}}
\newcommand{\kpc}{{\rm kpc}}
\newcommand{\pc}{{\rm pc}}
\newcommand{\phot}{{\rm phot}}
\newcommand{\s}{{\rm s}}
\newcommand{\yr}{{\rm yr}}

\newcommand{\NoteToEd}[1]{}

\newcommand{\masstable}{
    \begin{deluxetable}{ll}
    \tablewidth{0pt}
    \tablecaption{Fitted Model Parameters
    \label{masstable}}
    \tablehead{\colhead{Parameter} & \colhead{Value, with $3\sigma$ Error}}
    \startdata
    \protect\ion{Fe}{1} mass & $0.013^{+0.010}_{-0.005} \ \Msun$ \nl
    \protect\ion{Ca}{2} mass & $0.005^{+0.016}_{-0.002} \ \Msun$ \nl
    \protect\ion{Ca}{1} mass & $(2.9^{+2.4}_{-0.6}) \times 10^{-4} \ \Msun$ \nl
    \protect\ion{Fe}{1} central density & $0.0031^{+.0024}_{-.0011} \  \cm^{-3}$ 
\nl
    \protect\ion{Ca}{2} central density & $0.0016^{+0.0072}_{-0.0008} \  
\cm^{-3}$ \nl
    \protect\ion{Ca}{1} central density & $(9.7^{+7.3}_{-3.0}) \times 10^{-5} \  
\cm^{-3}$ \nl
    Maximum expansion velocity & $13100^{+1500}_{-1400} \  \kms$ \nl
    Foreground starlight fraction & $0.21^{+0.06}_{-0.12}$ \nl
    Aperture offset from center & $0.05^{+0.18}_{-0.05} \ \arcsecs$ \nl
    \enddata
    \end{deluxetable}
}

\newcommand{\theorytable}{
    \begin{deluxetable}{lcll}
    \tablecolumns{4}
    \tablewidth{0pt}
    \tablecaption{Fe and Ca Masses in SN~Ia Models
    \label{theorytable}}
    \tablehead{
	\colhead{Model} & \colhead{Fe} & \colhead{Ca} & \colhead{ref} \\
	\colhead{} & \colhead{(\Msun)} & \colhead{(\Msun)} & \colhead{}
    }
    \startdata
	\multicolumn{4}{c}{Subluminous SN~Ia} \\
	\hline
	PDD1c\tablenotemark{a} & 0.12 & 0.017 & H98 \\
	HeD6 & 0.18 & 0.011 & HK96, H98 \\
	Model 1 & 0.29 & 0.022 & WW94 \\
	\hline
	\multicolumn{4}{c}{Normal SN~Ia} \\
	\hline
	NCD6A & 0.56 & 0.0083 & W97 \\
	WDD2 & 0.70 & 0.035 & N97 \\
	DD21c & 0.73 & 0.040 & HWT98 \\
	W7 & 0.77 & 0.041 & NTY84
	\tablenotetext{a}{
	    PDD1c is an updated version of model PDD5 of
	    H\"{o}flich, Khokhlov \& Wheeler (1995).
	}
	\tablerefs{
	    (H98) H\"{o}flich 1998, private communication;
	    (HK96) H\"{o}flich \& Khokhlov 1996;
	    (HWT98) H\"{o}flich, Wheeler \& Thielemann 1998;
	    (N97) Nomoto et al.\ 1997;
	    (NTY84) Nomoto, Thielemann, \& Yokoi 1984;
	    (WW94) Woosley \& Weaver 1994;
	    (W97) Woosley 1997.
	}
    \enddata
    \end{deluxetable}
}

\newcommand{\image}{
    \begin{figure*}
    \begin{center}
    \leavevmode
    \epsfbox[94 41 496 670]{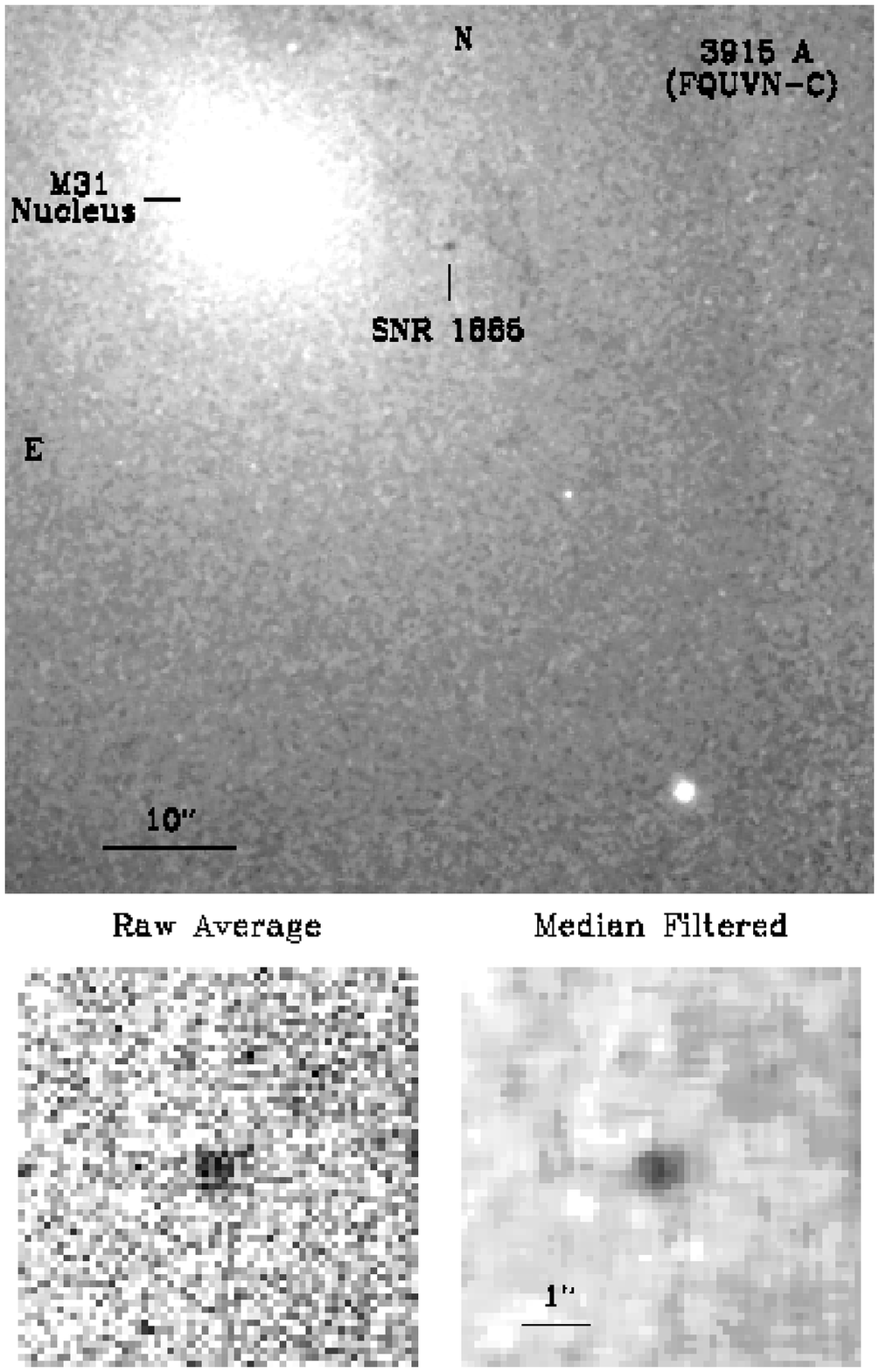}
    \end{center}
    \caption[1]{\small
    \imagecaption
    }
    \end{figure*}
}
\newcommand{\imagecaption}{
WFPC2 image of SNR~1885 in the bulge of M31.
\label{image}
}

\newcommand{\spec}{
    \epsfxsize=3.5in
    \epsfbox{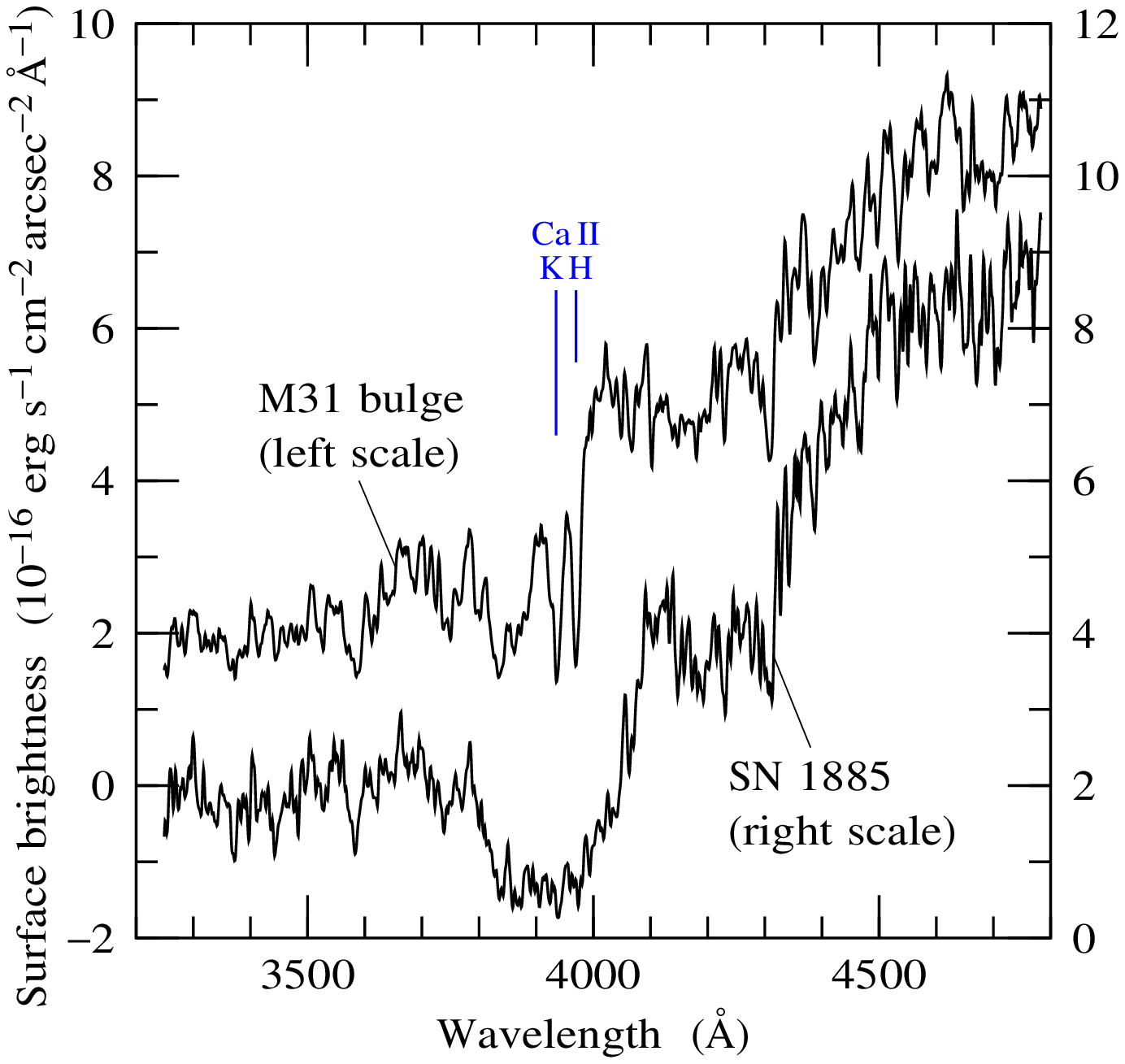}
    \caption[1]{\small
    \specfigcaption
    }
}
\newcommand{\specfigcaption}{
Coadded G400H FOS spectra of the bulge of M31,
and of the SNR~1885 absorption spot.
The spectrum of the bulge has been normalized to the same continuum level
as that of SNR~1885, and is offset upward by
$2 \times 10^{-16} \, \erg \, \s^{-1} \cm^{-2} \arcsecs^{-2} {\rm \AA}^{-1}$
for clarity.
Vacuum wavelengths have been transformed to the rest frame of M31.
Both spectra have been smoothed with a Gaussian of dispersion $120 \, \kms$.
\label{spec}
}

\newcommand{\fit}{
    \epsfxsize=3.4in
    \epsfbox{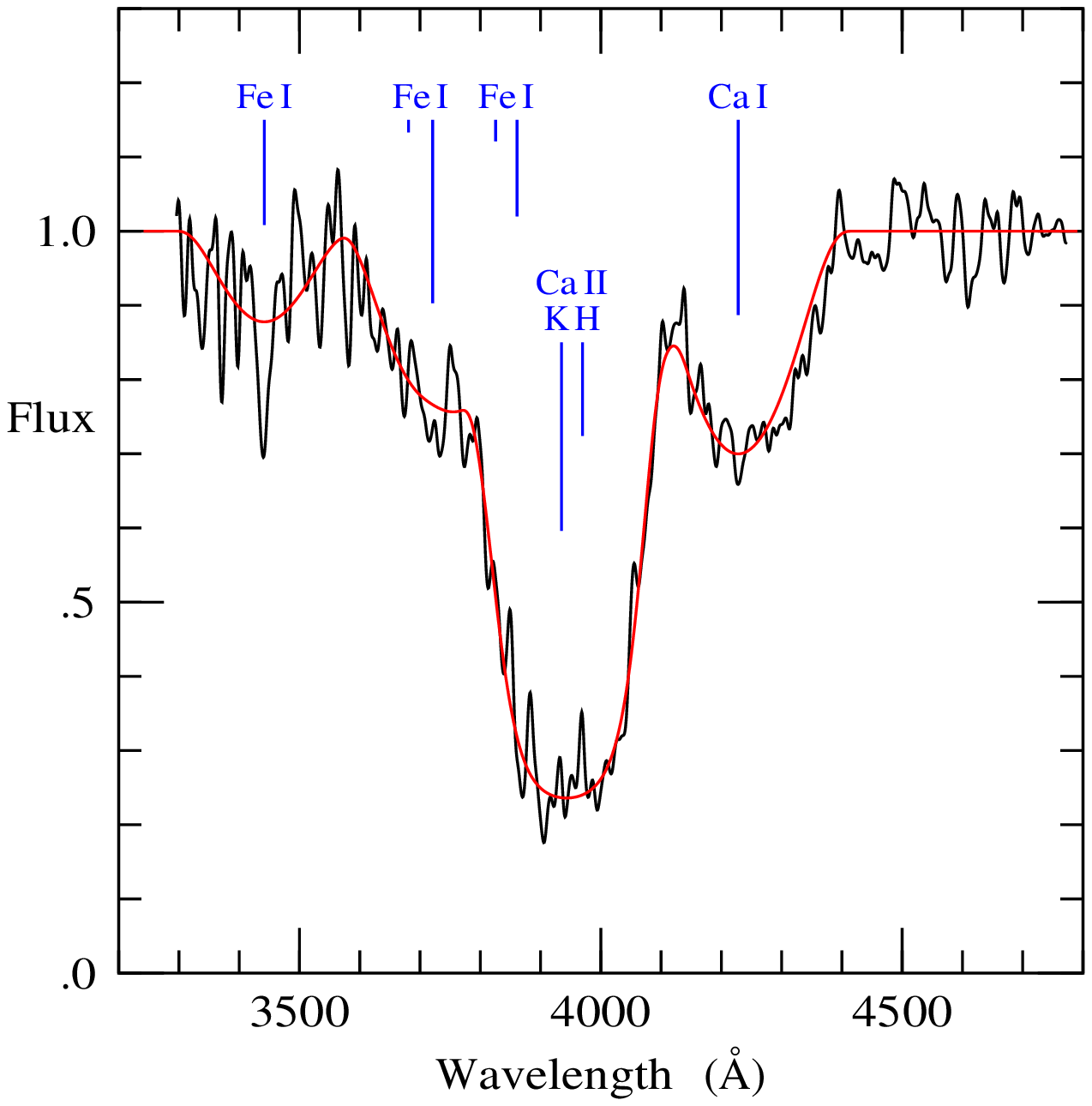}
    \small\caption[1]{\small
    \fitfigcaption
    }
}
\newcommand{\fitfigcaption}{
Best fit model spectrum (smooth line) compared to the
ratio of the FOS spectrum of SNR~1885 to that of the M31 bulge.
Both SNR~1885 and bulge spectra were smoothed with a Gaussian of
dispersion $300 \, \kms$ before taking their ratio.
Vertical lines, with lengths proportional to relative oscillator strengths
for that ion, mark wavelengths of lines included in the model.
All three ion species are assumed to have the same spherically symmetric,
bell-shaped density profile, equation (\protect\ref{n}).
Notice that the observed \ion{Ca}{1} absorption appears redshifted by
$\approx 1100 \, \kms$ compared to the model.
The residual spikes at H \& K at the bottom of the \ion{Ca}{2} aborption
trough are consistent with being noise.
\label{fit}
}

\newcommand{\specfitfigs}{
    \begin{figure*}[tb]
    \parbox[t]{3.5in}{
	\spec
    }
    \hfill
    \parbox[t]{3.5in}{
	\fit
    }
    \end{figure*}
}

\newcommand{\specuvfig}{
    \begin{figure*}[tb]
    \begin{center}
    \leavevmode
	\specuv
    \end{center}
    \end{figure*}
}
\newcommand{\specuv}{
    \epsfxsize=7in
    \epsfbox{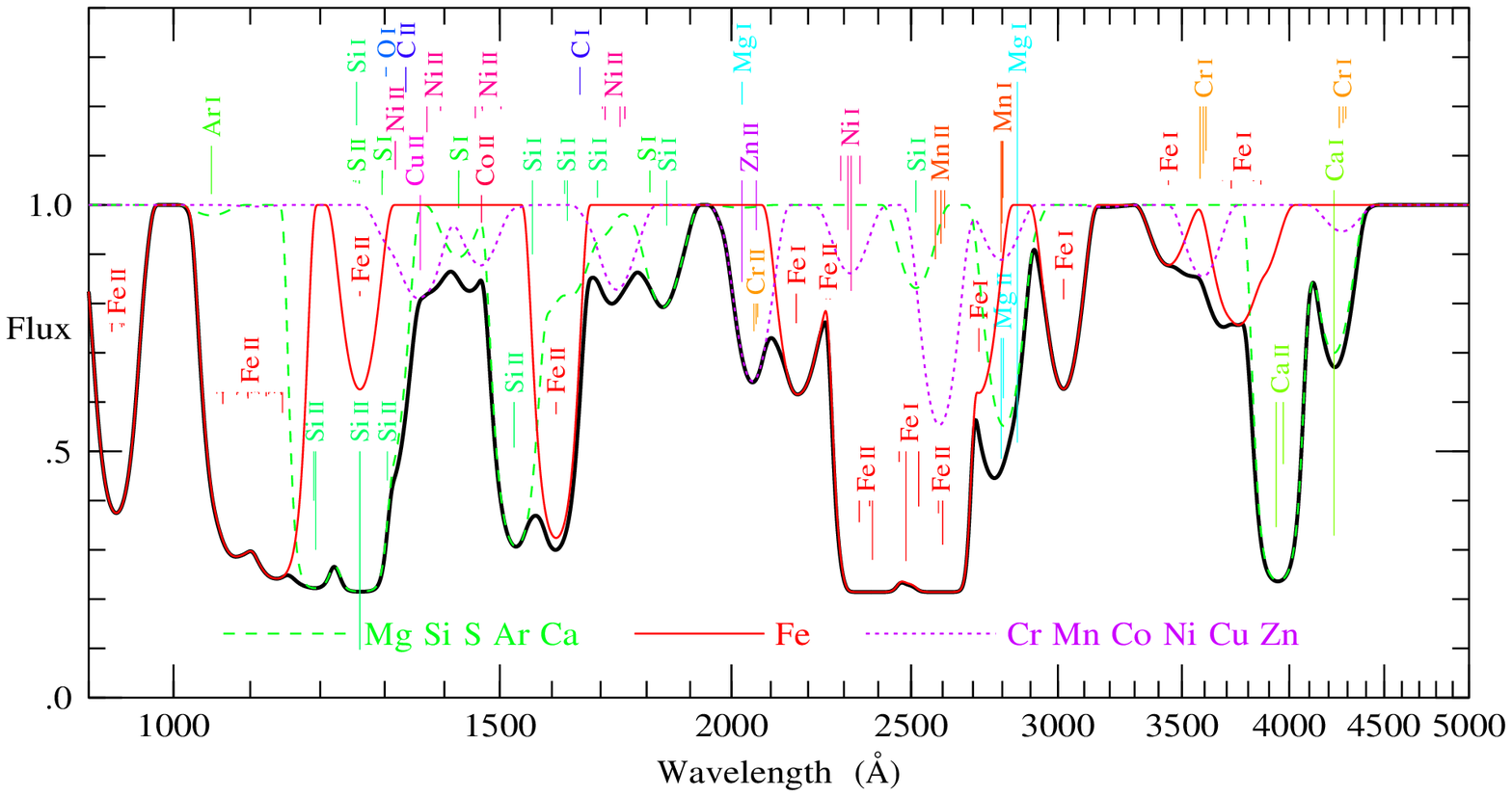}
    \small\caption[1]{\small
    \specuvfigcaption
    }
}
\newcommand{\specuvfigcaption}{
Predicted UV-optical absorption spectrum of SN~1885.
Vertical lines, with lengths proportional to oscillator strengths,
mark wavelengths of significant lines included in the model.
\label{specuv}
}

\begin{document}

\title{{\it HST\/} Images and Spectra of the Remnant of SN~1885 in M31}

\author{Robert A. Fesen \& Christopher L. Gerardy}
\affil{6127 Wilder Laboratory, Physics \& Astronomy Department
	Dartmouth College, Hanover, NH 03755
	\\ fesen@snr.Dartmouth.EDU, Christopher.L.Gerardy@Dartmouth.EDU
	\\ and }

\author{Kevin M. McLin  \& Andrew J. S. Hamilton}
\affil{JILA \& Dept. of Astrophysical \& Planetary Sciences, U. Colorado, 
	Boulder, CO 80309
	\\ mcl@casa.Colorado.EDU, Andrew.Hamilton@Colorado.EDU
	}

\begin{abstract}

Near UV {\it HST\/} images of the remnant of SN~1885 (S~And) in M31 show 
a $0\farcs 70 \pm 0\farcs 05$ diameter absorption disk silhouetted against 
M31's central bulge, at SN~1885's historically reported position.
The disk's size corresponds to a linear diameter of $2.5 \pm0.4 \, \pc$
at a distance of 725 $\pm 70$ kpc,
implying an average expansion velocity of $11000 \pm2000 \, \kms$
over 110 years.
Low-dispersion FOS spectra over $3200$--$4800 \, {\rm \AA}$
reveal that the absorption arises principally
from \ion{Ca}{2} H \& K
(equivalent width $\simeq  215 \, {\rm \AA}$),
with weaker absorption features of \ion{Ca}{1} $4227 \, {\rm \AA}$
and \ion{Fe}{1} $3720 \, {\rm \AA}$.
The flux at \ion{Ca}{2} line center indicates a foreground starlight
fraction of 0.21,
which places SNR~1885 some $64 \, \pc$ to the near side of
the midpoint of the M31 bulge,
comparable to its projected $55 \, \pc$ distance from the nucleus.
The absorption line profiles suggest an approximately spherically symmetric,
bell-shaped density distribution of supernova ejecta
freely expanding at up to $13100 \pm1500 \, \kms$.
We estimate \ion{Ca}{1}, \ion{Ca}{2}, and \ion{Fe}{1}  masses of 
$2.9^{+2.4}_{-0.6} \times 10^{-4} \, \Msun$,
$0.005^{+0.016}_{-0.002} \, \Msun$, 
and $0.013^{+0.010}_{-0.005} \, \Msun$ respectively.
If the ionization state of iron is similar to
the observed ionization state of calcium,
$M_\CaII/M_\CaI = 16^{+42}_{-5}$,
then the mass of \ion{Fe}{2} is
$ 0.21^{+0.74}_{-0.08} \, \Msun$, consistent with that expected for 
either normal or subluminous SN~Ia.

\end{abstract}

\keywords{
galaxies: individual (M31)
---
ISM: supernova remnants
---
stars: supernovae: individual (SN~1885)
---
ultraviolet: galaxies
---
ultraviolet: ISM
}

\section{Introduction}

SN~1885 (S~Andromedae) attained a peak $V$ magnitude of 5.85 in August 1885,
and remains the only supernova (SN) recorded in the Andromeda galaxy, M31.
Descriptions of its optical spectrum indicate a lack of hydrogen and
overall agreement with typical SN~Ia, except for the notable absence 
of \ion{Si}{2} $6150 \, {\rm \AA}$ absorption (\cite{DVC85},
hereafter dVC).
At M31's distance of $725 \pm70 \, \pc$ ($dM = 24.30 \pm 0.20$;
\cite{vdb91}; \cite{OG97} and references therein),
and assuming 0.23~mag $V$ extinction, SN~1885 had a peak brightness
of $M_V = -18.7$ (\cite{vdb94}), making it subluminous
compared to normal SN~Ia (\cite{Branch93}).
SN~1885 also exhibited an exceptionally
fast initial decline (2~mag in 12.5 days),
and was substantially redder near maximum ($B$$-$$V = +1.3$) than standard
SN~Ia events (dVC; \cite{Graham88}; \cite{CP88}; \cite{vdb94}).
These properties place SN~1885 at the subluminous end
of the range of Type~Ia supernovae,
similar to SN~1986G and SN~1991bg
(\cite{Fil92}; \cite{Lei93}; \cite{Hamuy96}).

The remnant of SN~1885 (SNR~1885) was detected
in 1988 by Fesen, Hamilton \& Saken (1989; hereafter FHS),
who used a near-UV filter
($\lambda_{\rm c}$ = $3900 \, {\rm \AA}$; FWHM = $200 \, {\rm \AA}$)
to image the remnant
as an absorption spot silhouetted against M31's central bulge.
FHS attributed the absorption to the resonance line of
\ion{Fe}{1} $3860 \, {\rm \AA}$,
consistent with the expected presence of a large mass of
iron in a SN~Ia.

In this paper, we present Hubble Space Telescope ({\it HST\/})
imaging and spectral data, which both resolve the
SNR~1885 absorption silhouette, and show that
\ion{Ca}{2} H \& K, not \ion{Fe}{1} $3860 \, {\rm \AA}$,
account for most of the remnant's near-UV absorption.

\image

\specfitfigs

\section{Observations}

\NoteToEd{
EDITOR: PLACE FIGURE 1 ON PAGE 2 OF THE PUBLISHED PAPER, TAKING UP THE ENTIRE 
PAGE.
}

\subsection{Images}

Motivated by the ground-based detection of SNR~1885,
we obtained {\it HST\/} ``on'' and ``off'' \ion{Fe}{1} $3860 \, {\rm \AA}$
absorption images using two of the WFPC2 Quad redshifted
[\ion{O}{2}] 3726, $3729 \, {\rm \AA}$ filters.
On 1995 August 7,
two sets of $2 \times 400 \, \s$ exposures,
centered near the location of SNR~1885,
were taken with the FQUVN-C filter
($3915 \pm 30 \, {\rm \AA}$; ``off''),
and two sets of $2 \times 500 \, \s$ exposures with the FQUVN-D filter
($3839 \pm 30 \, {\rm \AA}$; ``on'').
Each set was shifted 5 pixels in both $x$ and $y$ directions on the CCD,
and then unshifted and coadded to remove cosmic ray hits and hot pixels.

While the ``on-band'' $3839 \, {\rm \AA}$ images showed surprisingly 
little absorption, the supposed ``off-band'' $3915 \, {\rm \AA}$ images
revealed a dark spot at the site of SN~1885 (see Fig.~1, top).
The remnant's prominence in the $3915 \, {\rm \AA}$ image immediately
suggested that \ion{Ca}{2} 3934, $3968 \, {\rm \AA}$ (K \& H),
not \ion{Fe}{1} $3860 \, {\rm \AA}$,
was the cause of the absorption both here and
in FHS's ground-based images.

The dark spot contained an average 1.0 counts per pixel
in each $3915 \, {\rm \AA}$ exposure,
while adjacent regions of the bulge contained 2.2 counts per pixel.
These low counts result from the faintness of M31's bulge below
$4000 \, {\rm \AA}$, coupled with WFPC2's $0\farcs1$ pixel size.

From the stacked frames we measured a SNR diameter
of $0\farcs 70 \pm 0\farcs 05$, with no evidence for non-circularity
(see Fig.~1, bottom panels).
This angular size corresponds to a linear diameter of $2.5 \pm 0.4 \, \pc$
at M31's distance of $725 \pm 70 \, \kpc$,
which in turn implies an average expansion velocity of $11000 \pm 2000 \, \kms$
over $110 \, \yr$ (1885--1995).
This velocity is essentially the same as that seen in the absorption spectrum of
\ion{Ca}{2} (see below),
indicating that the ejecta are freely expanding.
The velocity is also similar to the velocities
seen in \ion{Ca}{2} absorption in SN~Ia spectra at early times
(\cite{Lei93}).

The coordinates of the SNR~1885 spot measured from the WFPC2 images are
$\alpha(2000) = 00^{\rm h} 42^{\rm m} 42\fs 89 \pm 0\fs 01$,
$\delta(2000) = +41^{\circ} 16' 05\farcs 0 \pm 0\farcs 1$.
This places SNR~1885 at
$15\farcs 04 \pm 0\farcs 10$ west and $4\farcs 1 \pm 0\farcs 10$ 
south of M31's nucleus,
in good agreement with the
offset distances of $15\farcs 4$~W and $3\farcs 95$~S
measured from historical plates (dVC).


\masstable

\subsection{FOS Spectra}

\NoteToEd{
EDITOR: PLACE FIGURES 2 \& 3 SIDE BY SIDE AT THE TOP OF PAGE 3 OF THE PUBLISHED 
PAPER.
}

Low-dispersion spectra of the SNR~1885 absorption spot
were obtained on 1995 November 14 using the G400H grating on
{\it HST\/}'s Faint Object Spectrograph (FOS),
covering 3240--$4780 \, {\rm \AA}$.
Total exposure time on target was $3000 \, \s$.
However, because of an offset star positioning error,
the spectrograph's 0\farcs43 diameter aperture
missed SNR~1885's 0\farcs70 disk following a blind offset slew.

The observations were re-attempted on 1996 December 30,
with integration time of $3760 \, \s$,
resulting in a successful detection of the absorption spectrum of SNR~1885.
High counts from a few hot detector channels were removed from 7 of the 31
accumulation
data groups, and replaced with interpolated averages of adjacent pixel values.

To provide a template background spectrum of starlight,
FOS spectra of the M31 bulge close to the bright nucleus
were also taken through a 0\farcs86 aperture,
a $630 \, \s$ exposure in 1995, and $380 \, \s$ in 1996.

Figure~\ref{spec} shows the FOS spectra of the M31 bulge,
and of the SNR~1885 absorption spot.
Wavelengths of all spectra have been transformed to the rest frame of M31
by correcting for M31's $-300 \, \kms$ blueshift,
plus an additional $-270 \, \kms$ blueshift that appears consistently
in all spectra and which we attribute to an FOS calibration error.

Comparison of the two bright bulge spectra with the fainter ones
obtained near SNR~1885's location (i.e.\ with the missed 1995 spectra)
reveals that the brighter bulge spectra are a tinge redder,
but otherwise not significantly different.
To produce the bulge spectrum shown in Figure~\ref{spec},
we blued the brighter spectra to the same color as the fainter SNR~1885 ones,
by multipling by $(\lambda/3935 \, {\rm \AA})^{-0.2}$,
before coadding all four spectra.

Dividing the spectrum of the SNR~1885 absorption spot by that of the bulge
effectively removes stellar features,
leaving a clean spectrum of the remnant,
as shown in Figure~\ref{fit}.
The SNR~1885 spectrum shows a broad, strong absorption feature
centered at $3944 \, {\rm \AA}$
(equivalent width $\simeq$ 215 \AA),
which is undoubtedly \ion{Ca}{2} 3934, $3968 \, {\rm \AA}$ (K \& H).
This absorption feature is flanked on its red side by a similarly broad
but weaker feature centered at $4244 \, {\rm \AA}$,
and on its blue side by a broad depression around $3700 \, {\rm \AA}$.
We identify the absorption feature to the red as
the strong $4227 \, {\rm \AA}$ resonance line of \ion{Ca}{1} 
(oscillator strength $f = 1.753$; \cite{Morton91}),
even though the observed line center is
redshifted by $\simeq 1100 \, \kms$,
and the depression to the blue as
\ion{Fe}{1} $3720 \, {\rm \AA}$, the strongest \ion{Fe}{1} resonance line
($f = 0.0412$) in the 3200--$4800 \, {\rm \AA}$ region.
No plausible and consistent alternative identifications 
(such as \ion{Cr}{1}) were found for these features.

\theorytable

\section{Analysis}

\NoteToEd{
EDITOR: PLACE TABLES 1 \& 2 AT APPROPRIATE PLACES ON PAGES 4 OR POSSIBLY 5 OF 
THE PUBLISHED PAPER.
}

\subsection{Model Fitting}
\label{model}

The {\it HST\/} image and spectrum of SNR~1885 provide information 
about the mass and density distribution of the supernova ejecta.
The spherical symmetry of the absorption image,
along with the approximate central symmetry of the
\ion{Ca}{2} and \ion{Ca}{1} absorption line profiles
(setting aside for the moment the $\simeq 1100 \, \kms$
redshift of \ion{Ca}{1}),
suggest that the intrinsic density distribution is
approximately spherically symmetric. 
Further, all three ion species
\ion{Ca}{2}, \ion{Ca}{1}, and \ion{Fe}{1}
appear to have comparably broad absorption profiles,
suggesting that their density distributions may be similar.
We therefore adopt the simplifying assumptions
(a) that the density distribution is spherically symmetric,
and (b) that the density distribution is the same
for all three ion species.

The likelihood that the strong \ion{Ca}{2} H and K absorption
is partially saturated at its center,
and the unknown level of starlight to the foreground of SNR~1885,
make interpretation of the central parts of the
\ion{Ca}{2} feature ambiguous.
The \ion{Ca}{1} $4227 \, {\rm \AA}$ feature is clearer in this respect,
since its breadth and shallowness suggest that it is optically thin,
and foreground starlight makes only a relatively small contribution.
The shape of the \ion{Ca}{1} $4227 \, {\rm \AA}$ absorption profile 
indicates that the intrinsic density distribution is centrally concentrated.
Trial fits to the \ion{Ca}{1} profile
assuming the FOS aperture was precisely centered on the remnant
indicate an approximately quadratic density distribution,
$n(v) \propto 1 - (v/v_{\max})^2$
with maximum velocity $v_{\max} \approx 11000 \, \kms$.

Applying the same quadratic density distribution to the \ion{Ca}{2} feature
leads to a maximum velocity some $1000 \, \kms$ or so larger.
This could mean that the \ion{Ca}{2} is more extended than \ion{Ca}{1},
or it could mean that the Ca density distribution has wings.
Both possibilities are plausible; we adopt the latter.

The \ion{Ca}{1} and \ion{Ca}{2} absorption profiles together fit nicely
to a bell-shaped intrinsic density distribution, a quartic
\be
\label{n}
  n(v) \propto \left[ 1 - (v/v_{\max})^2 \right]^2
  \quad (v \leq v_{\max})
  \ ,
\ee
and the same distribution works well also for \ion{Fe}{1}.
Again, in arriving at this density distribution,
we assumed that the FOS aperture was accurately centered on the remnant.

Figure~\ref{fit} shows the best fit model assuming the bell-shaped
density distribution of equation~(\ref{n}),
and Table~1 gives the fitted parameters of the model.
We choose to quote $3\sigma$ errors on the parameters,
in part to allow for uncertainty in the choice of model.

Much of the quoted uncertainties listed in Table~1 derive
from uncertainties in two quantities,
the fraction of starlight to the foreground of SNR~1885,
and the offset of the FOS aperture from the precise center of the remnant.
Increasing the fraction of foreground starlight from its best fit value
of 0.21 increases the quantity of \ion{Ca}{2} required
to maintain the observed depth of the absorption,
making the central parts of the \ion{Ca}{2} profile more optically thick,
and flattening the bottom of the profile.
Similarly, mis-centering the $0\farcs 43$ FOS aperture
within the $0\farcs70$ remnant
increases the amount of \ion{Fe}{1}, \ion{Ca}{2}, and \ion{Ca}{1}
required to maintain the observed depth of absorption.

The foreground starlight fraction can be used to infer
the line-of-sight distance between SNR~1885
and the midpoint of the bulge of M31,
given a model of the 3-dimensional distribution of starlight in M31.
The surface brightness profile of the bulge measured from the WFPC2 image
shows a cusp-like core within $\sim 2''$ radius, but outside this
the surface brightness fits a modified Hubble law
$\Sigma(r) \propto 1/[1 + (r/21'')^2]$
(cf.\ \cite{BT87}, p.\ 230).
For a foreground starlight fraction of $0.21^{+0.06}_{-0.12}$,
and at M31's $725 \, \kpc$ distance,
this modified Hubble model places SNR~1885 at $64^{+69}_{-16} \, \pc$
to the near side of the bulge midpoint,
comparable to its transverse distance of $55 \, \pc$ from the center of M31.

\specuvfig

\subsection{Masses}
\label{mass}

The dominant element in SNe~Ia is thought to be iron.
As discussed in \S\ref{photoionization} below,
the photoionization lifetime of \ion{Fe}{1} in SNR~1885 is rather short.
This,
along with the observed ionization state
$M_\CaII/M_\CaI = 16^{+42}_{-5}$
of calcium,
suggests that iron in SNR~1885 is probably mostly Fe~II
at the present time.
The fact that the ejecta are still freely expanding out to $13000 \, \kms$
indicates that to date they have swept up little interstellar gas,
so there is little shocked gas emitting the hard UV and x-ray radiation
capable of photoionizing iron or calcium to doubly-ionized or higher
ionization stages
(cf.\ Hamilton \& Fesen 1988).
Thus, there is probably little iron or calcium in ionization stages
higher than singly-ionized.

The mass of \ion{Fe}{2} can be estimated from 
$M_\FeII = x \, M_\FeI M_\CaII/M_\CaI$
where $x \equiv (M_\FeII/M_\FeI)\discretionary{}{}{}/(M_\CaII/M_\CaI)$
is the ionization state of iron relative to that of calcium.
This yields 
\be
  M_\Fe \approx
  M_\FeII = x \  0.21^{+0.74}_{-0.08} \  \Msun
  \ .
\ee
Since \ion{Fe}{1} has a somewhat higher ionization potential than \ion{Ca}{1}
($7.9 \, \eV$ vs.\ $6.1 \, \eV$),
the ionization state of Fe is likely to be somewhat
lower than that of Ca,
that is, $x$ is expected to be a number somewhat less than~1.

For comparison,
Table~\ref{theorytable} lists the mass of iron and calcium
predicted in a selection of model SNe~Ia taken from recent literature.
A mass of iron $\simeq 0.2 \, \Msun$ in SNR~1885
is more consistent with models of subluminous than normal SN~Ia.
However, the large uncertainty in the estimate
means that a higher Fe mass,
consistent with a normal SN~Ia, cannot be ruled out.
The mass of calcium $\simeq 0.005 \, \Msun$ in SNR~1885
(Table~\ref{masstable})
is also more consistent, on the whole,
with the lower calcium masses predicted by subluminous SN~Ia models.

The ratio of iron to calcium in SNR~1885 is
$M_\FeII/M_\CaII = x \, M_\FeI/M_\CaI = x\, 44^{+6}_{-9}$.
For comparison,
the solar system (meteoritic) ratio of calcium to iron is
$M_\Fe/M_\Ca \approx 15$
(\cite{AG89}; \cite{Hannaford92}).

\subsection{Predicted UV spectrum}
\label{uv}

\NoteToEd{
EDITOR: PLACE FIGURE 4 AT THE TOP OF PAGE 5 OR 6 OF THE PUBLISHED PAPER.
}

Figure~\ref{specuv}
shows the UV-optical absorption spectrum of SNR~1885
predicted by the model of \S\ref{model}.
The optical part of the spectrum is
(aside from the \ion{Cr}{1} lines; see below)
the same as the best fit spectrum shown in Figure~\ref{fit}.
The predicted spectrum includes absorption from
all expected non-negligible resonance lines
with wavelengths above $912 \, {\rm \AA}$, a total of 122 lines,
extracted from the list of Morton (1991).
The model includes
neutral and singly-ionized species of
C, O, Mg, Al, Si, S, Ar, Ca,
and iron-group elements with ionic charges from 22 to 30,
namely Ti, V, Cr, Mn, Fe, Co, Ni, Cu, and Zn.
The abundances of the optically observed ions \ion{Ca}{1}, \ion{Ca}{2},
and \ion{Fe}{1} were fixed at their best fit values.
Abundances of other elements, relative to Fe,
were taken from the normal SN~Ia model DD21c of
H\"{o}flich, Wheeler \& Thielemann (1998).
The ratio of neutrals to singly-ionized ions was set at 1:10
in all cases except Ca.

The model spectrum does not include continuum opacity.
Aside from possible dust,
probably the only important current source of continuum opacity is \ion{Fe}{1}.
The photoionization cross-section of \ion{Fe}{1}
(Lombardi, Smith \& Parkinson 1978; Hansen et al.\ 1977)
averages about $3 \times 10^{-18} \, \cm^2$
between threshold ($1569 \, {\rm \AA}$)
and Lyman\ $\alpha$ ($1216 \, {\rm \AA}$),
with a broad peak from many resonances between
$1400 \, {\rm \AA}$ and $1250 \, {\rm \AA}$.
At an \ion{Fe}{1} column density of
$3 \times 10^{16} \, \cm^{-2}$ through the center of the remnant,
this implies an average continuum optical depth of about $0.1$.
This continuum opacity is not included in the spectrum shown in
Figure~\ref{specuv},
in part because the available cross-section data, referenced above,
are not precise enough for the purpose.


The predicted UV spectrum shows broad, deep chasms of absorption
dominated by \ion{Fe}{2} and \ion{Si}{2}.
The strongest of these lines are optically thick and heavily blended,
so the weaker
\ion{Fe}{2} $1608 \, {\rm \AA}$
and
\ion{Si}{2} $1527 \, {\rm \AA}$
lines will be important in constraining abundances reliably
from future observations.

Other notable features of the model spectrum include the following:
The clearest line of \ion{Fe}{1} is $3021 \, {\rm \AA}$,
and of \ion{Si}{1} is $1846 \, {\rm \AA}$.
The least contaminated view of \ion{Ni}{2} is the complex
around $1730 \, {\rm \AA}$,
the strongest lines there being
\ion{Ni}{2} $1710 \, {\rm \AA}$ and $1742$, $1752 \, {\rm \AA}$.
Lines of \ion{Co}{2} $1466 \, {\rm \AA}$
and \ion{Cu}{2} $1359 \, {\rm \AA}$
appear in the window around $1400 \, {\rm \AA}$,
somewhat blended with \ion{S}{1} $1425 \, {\rm \AA}$
and several lines of \ion{Ni}{2}.
There is a fairly isolated feature consisting of
\ion{Cr}{2} $2057 \, {\rm \AA}$
and \ion{Zn}{2} 2026, $2062 \, {\rm \AA}$.
The strongest lines of C and O, namely
\ion{C}{1} $1657 \, {\rm \AA}$,
\ion{C}{2} $1334 \, {\rm \AA}$,
and \ion{O1}{1} $1302 \, {\rm \AA}$,
are marked in Figure~\ref{specuv},
but produce very little absorption at the adopted abundances.
Lines of Ti and V are included in the model,
but produce negligible absorption.

The model predicts three times more \ion{Cr}{1}
than the observed upper limit from
\ion{Cr}{1} 3579, 3594, $3606 \, {\rm \AA}$ and
\ion{Cr}{1} 4255, 4275, $4290 \, {\rm \AA}$.
We have chosen to include \ion{Cr}{1} at the predicted level
in Figure~\ref{specuv},
in part to emphasize the fact that the observed optical spectrum
already places an interesting upper limit on the strength of \ion{Cr}{1} lines.
The relative weakness of \ion{Cr}{1} is not much of a puzzle,
since the photoionization lifetime of \ion{Cr}{1} is even shorter
than that of \ion{Ca}{1} or \ion{Fe}{1} (cf.\ \S\ref{photoionization}),
so the abundance of \ion{Cr}{1} relative to \ion{Cr}{2}
could well be less than the 1:10 ratio assumed in the model.

The predicted UV spectrum of SNR~1885 is stunningly rich.
Unfortunately, the low surface brightness of UV starlight from the bulge of M31
means that the S/N ratio attainable with the Space Telescope Imaging 
Spectrograph (STIS)
on {\it HST} is marginal, even in a fairly long exposure.

\subsection{Photoionization}
\label{photoionization}

Ejecta in SNR~1885 should be partially photoionized 
by UV starlight from the M31 bulge.
Figure~\ref{specuv} indicates that the ejecta are
likely to be optically thick at some UV wavelengths,
and optically thin at others. For simplicity, we consider
what happens if the ejecta are optically thin at all wavelengths.

In the limit where the ejecta are optically thin,
and in the approximation that starlight from the bulge is spherically
symmetric about the center of M31,
the flux $F_\lambda$ seen by SNR~1885 at wavelength $\lambda$,
integrated over all directions,
is related to the dereddened surface brightness $\Sigma_\lambda(r)$
at transverse distance $r$ from the nucleus of M31,
as observed by us on Earth, by
\be
\label{F}
  F_\lambda
  = \int_0^{\pi/2} \Sigma_\lambda(a \sin\theta) \, 2\pi \sin\theta \, d\theta
\ee
where $a$ is the 3-dimensional distance between SNR~1885 and the nucleus.

Equation~(\ref{F})
permits an estimate of the lifetime
of ions in SNR~1885 exposed to photoionization by UV starlight,
in the optically thin limit.
We measured the surface brightness $\Sigma(r)$
from the WFPC2 image, averaged in circular annuli about the center of M31,
and we assumed that the surface brightness distribution is the same at all
wavelengths.
We then took the UV spectrum of the bulge
from the {\it IUE\/} data of Burstein (1988) down to $1225 \, {\rm \AA}$,
and from the {\it HUT\/} data of Ferguson \& Davidsen (1993)
down to $912 \, {\rm \AA}$,
dereddened using the Cardelli et al.\ (1989) extinction curve
with color excess $E_{B-V} = 0.11$ (\cite{FD93})
and $R \equiv A_V / E_{B-V} = 3.1$.
We normalized the surface brightnesses to the bulge spectrum
shown in Figure~\ref{spec}.
Finally,
we integrated over photoionization cross-sections $\sigma_\lambda$ from
Verner et al.\ (1996),
to derive photoionization lifetimes $t_\phot$, given by
$t_\phot^{-1} = \int \sigma_\lambda F_\lambda (\lambda/h c) \, d\lambda$.


The result is that
the lifetimes of \ion{Ca}{1} and \ion{Fe}{1} ions
in SNR~1885 exposed to UV starlight from the bulge of M31 are,
in the optically thin limit,
$8^{+10}_{-1} \, \yr$ and $10^{+12}_{-1} \, \yr$ respectively.
Quoted uncertainties here include only those arising from
($3\sigma$) uncertainty in the 3-dimensional distance
$a = 84^{+60}_{-11} \, \pc$
of SNR~1885 from the nucleus of M31,
not uncertainties in the extinction,
the photoionization spectrum or surface brightness distribution,
or in photoionization cross-sections,
which could plausibly lead to an additional factor of two uncertainty.

Recombination is negligible.
While expansion acts as an efficient refrigerator,
the competition between adiabatic cooling and photoionization heating
probably leads to T $\sim 10^{3} - 10^{4}$ K.
At any temperature exceeding  $1 \, {\rm K}$,
the sum of radiative and dielectronic recombination rates is no more
than $\sim 3 \times 10^{-11} \, \cm^3 \, \s^{-1}$
for either \ion{Ca}{1} or \ion{Fe}{1}
(Arnaud \& Rothenflug 1985).
The electron density at the densest point,
the center of the remnant, is probably no more than
$0.1 \, \cm^{-3}$, assuming that \ion{Fe}{2} is the most abundant ion
and that \ion{Fe}{2}/\ion{Fe}{1} is no more than $\sim 30$.
These estimates imply that recombination times exceed
$10^4 \, \yr$,
two orders of magnitude longer than the age of the remnant.

The photoionization times of $8 \, \yr$ for \ion{Ca}{1}
and $10 \, \yr$ for \ion{Fe}{1} are for optically thin ejecta,
whereas Figure~\ref{specuv} shows that the ejecta are likely to be
optically thick in broad bands of the ultraviolet.
A rough estimate based on the results of Figure~\ref{specuv}
suggests that the photoionization times should be increased
by a factor of perhaps 2 to allow for optical depth effects,
implying photoionization times $\sim 20 \, \yr$.

Such short photoionization lifetimes offer a natural explanation
of the fact that calcium in SNR~1885 is observed to be mostly singly-ionized,
$M_\CaII/M_\CaI = 16^{+42}_{-5}$.
Other elements in the ejecta can similarly be expected to be mostly
singly-ionized, as assumed in the model spectrum of Figure~\ref{specuv}.

At the same time, the photoionization lifetimes of \ion{Ca}{1} and \ion{Fe}{1}
are short enough as to raise the question of why one sees these ions at all?
One plausible explanation is that the ejecta are in the process of making
a transition from being optically thick to optically thin to
photoionizing radiation,
and that the ejecta are currently
undergoing a period of rapid photoionization out of the neutral state.
We pointed out in \S\ref{uv}
that the continuum optical depth of \ion{Fe}{1} through the center
of the remnant is currently of order
$0.1$ in the UV at wavelengths shorter than the $1569 \, {\rm \AA}$
ionization threshold of \ion{Fe}{1}.
We have also argued that there is currently about 10 times
as much \ion{Fe}{2} as \ion{Fe}{1}. If all of this iron were \ion{Fe}{1} rather 
than \ion{Fe}{2}, 
then the continuum optical depth of the \ion{Fe}{1} would be increased to about 
unity.
Consequently the \ion{Fe}{1} would then be in effect self-shielded,
inhibiting its photoionization.
Since column densities in the expanding ejecta
are decreasing with time $t$ as $t^{-2}$,
optical depths would have been higher in the past,
thus providing better self-shielding.
Thus the picture that emerges is that the ejecta
may be currently in the process of thinning out sufficiently to 
undergoing a period of rapid photoionization out of the neutral state.
Hamilton \& Fesen (1991) previously arrived at the same conclusion.

It remains to be seen whether this picture can account quantitatively
for the fact that \ion{Ca}{1} as well as \ion{Fe}{1} remain observable
in spite of their rather short current photoionization lifetimes.
Whatever the case, this self-shielding scenario makes a definite observational 
prediction,
that there should be an observable reduction in absorption from \ion{Ca}{1}
and \ion{Fe}{1} over a timescale of a decade or two.


\subsection{The puzzle of redshifted \ion{{\rm Ca}}{1}}

A final problem, possibly related to the problem of the 
short photoionization lifetime
of \ion{Ca}{1},
is that the observed
\ion{Ca}{1} $4227 \, {\rm \AA}$ absorption feature appears redshifted by
$\simeq 1100 \, \kms$,
as is evident in Figure~\ref{fit}.
There is also a hint of an abrupt blue edge
at $4142 \, {\rm \AA}$ on the \ion{Ca}{1} feature,
$-6200 \, \kms$ blueward of $4228 \, {\rm \AA}$ (vacuum) line center.
The possible blue edge
is unlikely to be caused by a shock front
(cf.\ \cite{Hamilton97})
since no corresponding blue edge is observed on the \ion{Ca}{2} feature.

On the other hand, the  redshift of the \ion{Ca}{1} line profile
could be caused by photoionization,
with the abrupt blue edge a possible sign of a photoionization front.
Hamilton \& Fesen (1991)
argued that the remnant of SN~1885 should be receiving a one-sided tan,
being photoionized more on the side facing the nucleus of M31.
This would predict that \ion{Ca}{1} should if anything
appear blueshifted,
and any photoionization front should be on the red side,
opposite to what is observed.

The possibility that the apparent redshift is caused by
\ion{Cr}{1} 4255, 4275, $4290 \, {\rm \AA}$
absorption is excluded by the absence of the stronger
\ion{Cr}{1} 3579, 3594, $3606 \, {\rm \AA}$
lines (cf.\ Figs.~\ref{fit} and \ref{specuv}).

We can offer no compelling solution to this puzzle.
The observed symmetry of the \ion{Ca}{2} line
suggests that the supernova ejecta as a whole are not grossly aspherical.
Perhaps however the \ion{Ca}{1} is concentrated into optically thick lumps,
distributed unevenly through the remnant, with our FOS aperture detecting
more red than blueshifted clumps.
Perhaps an interstellar
dust cloud happens to intervene between SNR~1885 and the M31 nucleus,
allowing photoionization to proceed more rapidly on the near,
blueshifted, side.
J. M. Shull (1998, private communication) suggests the possibility
that Ca is locked into grains,  
and \ion{Ca}{1} is being liberated
by photo-evaporation preferentially on the side nearer the nucleus.

\section{Conclusions}

{\it HST\/} FOS spectra of the remnant of the apparently subluminous Type~Ia
supernova of 1885 (S~And) in M31
reveal the presence of broad absorption lines of 
\ion{Ca}{2} 3934, $3968 \, {\rm \AA}$ (K \& H),
\ion{Ca}{1} $4227 \, {\rm \AA}$,
and \ion{Fe}{1} 3720 \AA.
The absorption line profiles indicate a bell-shaped distribution of supernova 
ejecta expanding at up to $13100 \pm 1500$ km s$^{-1}$.

{\it HST\/} WFPC2 images of the remnant show an absorption spot
$0\farcs 70 \pm 0\farcs 05$ in diameter,
corresponding to a free expansion velocity of $11000 \pm 2000 \, \kms$
at the $725 \pm 70 \, \kpc$ distance of M31.
The agreement between the expansion velocities inferred from spectrum and image
demonstrates that the ejecta are freely expanding.
The presence of low ionization species in the ejecta,
\ion{Ca}{1}, \ion{Ca}{2}, and \ion{Fe}{1},
is consistent with the notion that the ejecta are freely expanding
and unshocked.

The flux at \ion{Ca}{2} line center indicates a foreground starlight
fraction of $0.21^{+0.06}_{-0.12}$,
which places SNR~1885 some $64^{+69}_{-16} \, \pc$ to the near side of
the midpoint of the M31 bulge,
comparable to its projected $55 \, \pc$ distance from the nucleus.

The presence of \ion{Ca}{1} and \ion{Ca}{2} lines is consistent both
with the appearance of calcium lines in the spectra of SN~Ia near maximum
light (\cite{Lei93}; Filippenko 1997),
and with the predictions of SN~Ia models.
The masses of \ion{Ca}{1}, \ion{Ca}{2}, and \ion{Fe}{1}
inferred from the absorption spectrum of SNR~1885
are $0.0003$, $0.005$, and $0.013 \, \Msun$ respectively
(Table~\ref{masstable}).
If the ionization state of iron is similar to that of calcium,
then the mass of \ion{Fe}{2} is
$\approx 0.21^{+0.74}_{-0.08} \, \Msun$.
These low masses are more consistent with SN~1885 having been a subluminous
event (Table~\ref{theorytable}), as suggested by the historical record,
but the large uncertainties do not exclude a normal SN~Ia.

We estimate that the lifetimes of \ion{Ca}{1} and \ion{Fe}{1}
against photoionization by UV starlight from the bulge of M31
are only $\sim 20 \, \yr$.
We have argued that the ejecta are in the process of making
a transition from being optically thick to optically thin to
photoionizing radiation,
and that the ejecta are currently
undergoing a period of rapid photoionization out of the neutral state.
If this is correct,
then there should be an observable reduction in the strength of the
\ion{Ca}{1} and \ion{Fe}{1} absorption lines over a timescale of a decade
or two.

We have pointed out, but are unable to explain in a natural way,
an apparent $\simeq 1100 \, \kms$
redward displacement of the \ion{Ca}{1} $4227 \, {\rm \AA}$ absorption feature
from its expected line center.

The model inferred from the optical spectrum of SNR~1885,
coupled with abundances expected in SN~Ia,
predicts a rich UV spectrum of absorption lines.
If a UV spectrum with adequate signal-to-noise ratio could be obtained,
then in principle it would be possible to infer the masses of
Mg, Si, Cr, Co, Ni, Cu, and Zn,
in addition to Ca and Fe.
Unfortunately the faintness of the bulge of M31 in the UV
means that the S/N ratio currently attainable with STIS on {\it HST} is 
marginal.

Subluminous events like SN~1885, once thought to be anomalous, 
have recently become better understood as part of a wider range of
Type~Ia supernovae
(\cite{Graham88}; \cite{Hamuy96}).
Unusually red and faint SNe~Ia like SN~1991bg have been 
interpreted as originating from lower mass progenitors 
(\cite{CP88}; \cite{Fil92};
\cite{Lei93}; \cite{Nugent97}).
Future observations of SNR~1885 in absorption
offer a potentially powerful tool for understanding not only SN~1885 itself,
but also the class of subluminous SNe~Ia in general.

\acknowledgments

We thank P. H\"{o}flich for communicating abundances from his Type Ia models,
and M. Shull and R. McCray for helpful conversations.
Support for this work was provided by NASA through grant number
GO-6125 from the Space Telescope Science Institute,
which is operated by AURA, Inc., under NASA contract NAS 5-26555.

\end{document}